\documentclass[conference]{IEEEtran}
\usepackage{algorithm}
\usepackage{array}
\usepackage[caption=false,font=normalsize,labelfont=sf,textfont=sf]{subfig}
\usepackage{url}
\usepackage{verbatim}
\usepackage{tikz}
\usepackage{circuitikz}
\usetikzlibrary{positioning}

\usepackage{cite}
\usepackage{amsmath,amssymb,amsfonts}
\usepackage{algorithmic}
\usepackage{graphicx}
\usepackage{textcomp}
\usepackage{xcolor}
\def\BibTeX{{\rm B\kern-.05em{\sc i\kern-.025em b}\kern-.08em
    T\kern-.1667em\lower.7ex\hbox{E}\kern-.125emX}}
    
\begin{document}

\title{Novel Data-Driven Indices for Early Detection and Quantification of Short-Term Voltage Instability from Voltage Trajectories}

\author{
\IEEEauthorblockN{Mohammad Almomani, Muhammad Sarwar, and Venkataramana Ajjarapu}
\IEEEauthorblockA{\textit{Department of Electrical \& Computer Engineering, Iowa State University, Ames, IA, USA} \\
Emails: \{mmomani, msarwar, vajjarap\}@iastate.edu}
}

\maketitle

\begin{abstract}

This paper presents a novel Short-Term Voltage Stability Index (STVSI), which leverages Lyapunov Exponent-based detection to assess and quantify short-term stability triggered by Over Excitation Limiters (OELs) or undamped oscillations in voltage. The proposed method is measurement-based and decomposes the voltage trajectory into two key components using Empirical Mode Decomposition (EMD): a residual part, which indicates delayed voltage recovery, and an oscillatory part, which captures oscillations. The residual component is critical, as it can detect activation of OELs in synchronous generators or Low Voltage Ride-Through (LVRT) relays in inverter-based resources, potentially leading to instability within the quasi-steady-state time frame. Meanwhile, the oscillatory component may indicate either a stable or unstable state in the short term. 
To accurately assess stability, STVSI employs an entropy-based metric to measure the proximity of the system to instability, with specific indices for short-term voltage stability based on oscillations and recovery. Simulations on the Nordic power system demonstrate that STVSI effectively identifies and categorizes voltage stability issues. Moreover, STVSI not only detects voltage stability conditions but also qualitatively assesses the extent of stability, providing a nuanced measure of stability.
\end{abstract}

\begin{IEEEkeywords}
Delayed Voltage Recovery, Short-term Voltage Stability Index, Lyapunov Exponent

\end{IEEEkeywords}

\section{Introduction}
\textcolor{black}{Voltage stability assessment is critical for ensuring the reliable operation of power systems, particularly during transient disturbances that push the system away from equilibrium. The Lyapunov exponent (LE) provides a robust method for analyzing nonlinear system dynamics \cite{LE1}, offering insights into the stability of system trajectories by quantifying their exponential convergence or divergence rates. Unlike traditional equilibrium-based methods, LE effectively addresses the nonlinear, transient nature of short-term voltage stability problems by extending the concept of eigenvalues to nonlinear systems. A system with a negative maximum LE is considered stable, as its trajectories converge, while a positive maximum LE indicates instability due to exponential trajectory divergence \cite{LE1}. However, while LE excels in stability assessment, its conventional application in power systems often focuses on binary classification, offering limited insight into the specific characteristics of voltage deviations during transient events \cite{TSA1,TSA2, VSA1, TSA3}.}

Conversely, the Kullback-Leibler (KL) divergence presents a quantitative measure that effectively captures the probability distribution of deviations by assessing how one probability distribution diverges from a reference distribution. Widely used in literature, the KL divergence measure \cite{KL1} has proven particularly useful for quantifying Fault-Induced Delayed Voltage Recovery (FIDVR), as it compares post-fault voltage deviations against a predefined reference, thus motivating its application in quantifying stability strength during transient disturbances.

EMD adds further value in this context as a powerful technique for breaking down complex, non-linear, and non-stationary signals into intrinsic mode functions (IMFs). This decomposition facilitates the analysis of sophisticated voltage behaviors in power systems, allowing the separation of trend and oscillatory components following disturbances. By applying EMD, we can examine these signal characteristics in finer detail, calculating the LE for each decomposed component to assess stability levels, offering a deeper understanding of system dynamics during FIDVR events.

Together, these methodologies inspire the development of a comprehensive stability index that combines the strengths of KL divergence, LE, and EMD. This integrated approach aims to create a more accurate, robust, and insightful index for short-term voltage stability assessment for the modern power system.
\subsection{Literature Review and Research Gap}
Previous studies, such as \cite{TSA1} and \cite{TSA2}, have applied LE to transient rotor angle stability by computing the system LE using data from Phasor Measurement Units (PMUs) and system models. In \cite{VSA1}, the authors employ PMU data in a model-free and data-driven approach to monitor short-term voltage stability, allowing online LE computation.
However, after a severe disturbance, the MLE may oscillate between positive and negative values for an extended period, making it challenging to determine the system’s stability based solely on observing a positive or negative MLE, as future fluctuations remain uncertain \cite{LE3}. Dealing with this gap, reference \cite{TSA3} uses a recursive least-squares algorithm to estimate the MLE from real-time rotor angle measurements, with careful selection of the Theiler window and initial estimation time step to ensure that the resulting MLE curves distinctly indicate different stability conditions. \textcolor{black}{By leveraging the MLE of phase angles, method in \cite{10823530} enhances the reliability of stability assessments during transient events.}
In cases of delayed voltage recovery with oscillations, the fluctuation between positive and negative MLE values differs significantly from that observed in rotor angle stability, a scenario not addressed in \cite{TSA3,10823530}. 

It's worth noting that the MLE method is widely used in PMU applications for real-time monitoring but it is also well-suited for Simulated Data Analysis (SDA). This approach can facilitate early termination of simulations and help identify early onset of instability. Moreover, MLE has not been utilized in the literature to distinguish between transient angle stability and short-term voltage stability, highlighting a potential direction for future research.

\subsection{Contribution of This Paper}

The major contributions of this work are given below:

\begin{itemize}
    \item This paper addresses the limitations of LE in cases involving oscillations during delayed voltage recovery, providing a refined approach to capture complex short-term voltage stability dynamics by observing voltage trajectories with oscillations.
    \item We propose to systematically identify the critical boundary operating point of the short-term voltage stability occurring because of the delayed voltage recovery when a generator is tripped because of OEL.
    \item Based on the critical operating point identified, we develop a stability assessment method that quantifies the degree of stability, moving beyond conventional binary classification (stable/unstable) to offer a more nuanced assessment of short-term voltage stability.
\end{itemize}

\textcolor{black}{In contrast to existing methods that primarily detect stability or instability as a binary state, the proposed approach offers a significant advancement by quantifying the degree of stability and providing actionable insights into the dynamic behavior of power systems. This is achieved through the development of the STVSI that decomposes voltage trajectories into residual and oscillatory components, enabling separate evaluations of delayed voltage recovery and oscillatory instability. Unlike traditional Lyapunov Exponent (LE)-based methods, which often face challenges with oscillatory behavior or provide binary detection, our approach introduces a more sensitive and robust measure that not only identifies instability earlier but also distinguishes between instability caused by slow voltage recovery and that caused by undamped oscillations. Section IV demonstrates the practical application of the proposed index through simulations on the Nordic system, where it consistently outperforms conventional methods by offering earlier warnings and more detailed stability insights, aiding system operators in proactive decision-making.}

The rest of this paper is organized as follows. In Section II, we provide background information on LE, EMD and KL divergence. The proposed index is detailed in Section III, while simulation results using the Nordic system are presented in Section IV. Finally, conclusions are summarized in Section V.

\section{Background}

\subsection {Lyapunov exponent (LE)}
The authors in \cite{LE2} proposed a measurement-based method for LE calculation, where the time-dependent deviation of voltage trajectories is described as:
\begin{equation}
\|\Delta v(t)\| = e^{\lambda (t)} \|\Delta v_0\|
\end{equation}
To account for measured voltage profiles, the authors in \cite{VSA1} approximate the LE over a finite data window as:
\begin{equation}
\lambda (k) = \frac{1}{k \cdot \Delta t} \ln \left( \frac{\left| v'((n_0 + k)\Delta t) \right|}{\left| v'(n_0 \Delta t) \right|} \right)
\label{eq:lamda}
\end{equation}

where $k$ represents the samples order in the data window, $\Delta t$ is the sampling interval, $n_0$ is the index of the first data point, and $v'(n_0\Delta t)$ is the time derivative of the voltage at the point $n_0\Delta t$ . 

The method proposed in \cite{VSA1} for LE calculation is sensitive to oscillations in the voltage profile, which may lead to misinterpretations of stability. Specifically, in the case of an oscillatory voltage profile with exponential recovery, the slope of the voltage trajectory at a future time point may be higher than the slope at the current time, even though the system is stable. For example, a simple voltage profile combining a damped oscillation with an exponential recovery, such as $v(t) = e^{-\alpha t}(1 + A \cos(\omega t))$, demonstrates that the oscillatory component can cause temporary increases in the slope, especially if A is large. The time derivative of such a profile, $v'(t) = -\alpha e^{-\alpha t}(1 + A \cos(\omega t)) - A \omega e^{-\alpha t}  \sin(\omega t)$, shows that the oscillatory term $\sin(\omega t)$ can significantly influence the slope over time, momentarily increasing it even when the system remains stable. This oscillation-induced fluctuation in the slope may cause the LE algorithm to overestimate instability \cite{LE3}.

\subsection{Empirical Mode Decomposition}

EMD is a powerful, adaptive signal processing technique introduced by N. E. Huang and colleagues in 1998 \cite{EMD8}. It is specifically designed to analyze non-linear and non-stationary time series data by decomposing the original signal into a set of IMFs and a residual. Each IMF represents a simple oscillatory mode, capturing different frequency components of the signal. This process is analogous to an adaptive wavelet transform but without requiring predefined basis functions. The flexibility and data-driven nature of EMD make it particularly suitable for our application.

The computational process of EMD involves several iterative steps \cite{wang2010intrinsic}.  The result of the EMD process is a decomposition of the original signal into a finite number of  IMFs, each representing a distinct oscillatory mode embedded in the data. The final residue captures the overall trend of the signal. Mathematically, the decomposition is represented as:
\[
v(t) = \sum_i \text{IMF}_i + R(t)
\]
where \(v(t)\) is the original signal, \(\text{IMF}_i\) represents the intrinsic mode functions, and \(R(t)\) is the final residual.

\subsection{KL Divergence Measure }

KL divergence, also known as relative entropy, is a fundamental concept in information theory used to quantify the difference between two probability distributions, \(P\) and \(Q\). Mathematically, KL divergence is defined as:
\begin{equation}
\centering
   D_{KL}(P||Q) = \sum_i P(i) \log \frac{P(i)}{Q(i)}
\label{eq:KL} 
\end{equation}
for discrete distributions and as an integral for continuous distributions. This measure is non-symmetric and always non-negative, with \(D_{KL}(P||Q) = 0\) if and only if \(P\) and \(Q\) are identical. In the context of FIDVR, KL divergence is used to quantify the deviation of post-fault voltage profiles from the ideal behavior, denoted as \(Q=P^{ref}\), providing a rigorous measure for detecting and analyzing FIDVR events \cite{KL1}. So, it is applied to find the statistical ``distance" between the observed voltage signal and the reference ideal signal \(P^{ref}\).

 \section{Proposed Short-Term Voltage Stability Index (STVSI)}

The proposed index combines two statistical measures to assess short-term voltage stability from post-fault voltage trajectory occurring because of two sources: 1) delayed voltage recovery (stability issue related to FIDVR happening because of OEL-based generator tripping), 2) instability occurring because of undamped oscillations in voltage. The FIDVR-related stability measure (recovery stability index: $D_{\text{KL}}^{r}$) evaluates the KL divergence between the distribution of the LE of the residual component and a reference signal, which is carefully chosen to be a shifted-reversed Gompertz distribution function. The residual, $R(t)$, represents the monotonical trend, and smaller KL values suggest better voltage recovery indicating stability

The oscillations-induced stability measure (oscillation stability index: $D_{\text{KL}}^{imf}$) assesses the stability of oscillatory components by calculating the KL divergence between the distribution of the LE of the IMFs  and a shifted-reversed Gompertz  distribution function. High KL values indicate ineffective damping, signaling potential stability issues.

The proposed stability indices are defined as
{\small
\begin{equation}
    \begin{split}
       D_{\text{KL}}^{r} =\log (R(t_0)) \times D_{\text{KL}}(P_1 \parallel {\sigma}(\gamma_1)) \\
       D_{\text{KL}}^{imf} = D_{\text{KL}}(P_2 \parallel {\sigma}(\gamma_2))
    \end{split}
\label{eq:proposed}
\end{equation}}

for recovery and oscillation-induced stability, respectively. 
Where:
\begin{itemize}
    \item \(P_1 \sim e^{\lambda_{R (t)}}= \left(\frac{|\dot{R}(i\Delta t+t_0)|}{|\dot{R}(t_0)|}\right)^{\frac{1}{i\Delta t}}\): Represents the distribution of  LE calculated from \(R(t)\).
    \item \(P_2 \sim e^{\lambda_{\text{IMFs}} (t)} = \left(\frac{\dot{\text{IMFs}}(i \Delta t+t_0)}{\dot{\text{IMFs}}(t_0)}\right)^{ \frac{1}{i \Delta t}}\): Expresses the distribution of LE derived from the IMFs.
    \item \(\sigma (\gamma)\): Denotes the shifted-reversed Gompertz distribution function with pdf \(e^{-e^{\gamma (x-0.5)}}, x>0\), serving as the probability distribution of the reference signal for both \(P_1\) and \(P_2\). 
\end{itemize}

For further analysis within the stability framework, it is crucial to establish two critical threshold values:  periodic oscillatory behavior at the boundary of instability for the critical value of oscillation stability index (\(D_{\text{critical}}^{\text{imf}}\)), and slow voltage recovery without oscillation at the verge of instability for the recovery stability index (\(D_{\text{critical}}^r\)). These thresholds are designed to define a critical stable case. When the measured value exceeds the threshold, it indicates instability; if it remains below, it indicates stability. The qualitative measure of the stability strength is given by the calculated value of both indices to detect and establish the degree of stability.

\paragraph{ Oscillation stability index threshold (\(D_{\text{critical}}^{\text{imf}}\))} this threshold represents the KL divergence between the distribution of LE for a fixed magnitude oscillation around the nominal voltage (1 pu) and the shifted-reversed Gompertz distribution function. This index can be calculated analytically from (\ref{eq:proposed}) and is equal to 1 (after considering the normalization of the Gompertz distribution function given in the region of operation). If \(D_{KL}^{\text{imf}}\) exceeds 1, the system is considered unstable; if it is less than 1, the system is stable. A smaller value indicates a more stable case. This formulation captures the desired stability threshold, where \(\gamma_2\) in (\ref{eq:proposed}) is a parameter indicating the sensitivity of the index. Mathematically, if \(\gamma_2 \rightarrow \infty\), the index functions as a binary classifier. Conversely, a very small value of \(\gamma_2\) (approaching zero) reduces the accuracy of the index. Therefore, achieving a suitable trade-off is crucial. In this study it is selected to be 1.
 
\paragraph{  Recovery-caused stability index threshold (\(D_{\text{critical}}^r\))} Slow recovery cases are typically identified through quasi-steady state analyses and may trigger protective limits such as the OEL in traditional synchronous generators, or LVRT capabilities in IBRs which may lead to a deficit in reactive power,  potentially leading to quasi-steady state instability. The critical signals defined in a quasi-steady state are used here to quantify this threshold. Proper tuning of \(\gamma_1 \) is crucial for accurately differentiating between stable and unstable states. Particularly, it is essential when multiple interconnected recovery profiles influence quasi-steady state stability. For instance, profiles that exhibit rapid but large voltage dip or slow but small voltage dip should both be considered when adjusting \(\gamma_1\).


\subsection{Step-by-Step Implementation}

\subsubsection{Step 1: Stability Assessment Using KL Divergence}

The first step consists of two sub-measures: recovery index and oscillation index. Both use KL divergence, but for different signals.

\paragraph{ Recovery Index}
For recovery-caused stability, we calculate the KL divergence between the probability distribution of the LE ( \( e^{\lambda_R (t)}\) ), calculated using (\ref{eq:lamda}), of the residual component and a shifted-reversed Gompertz distribution function. let \(\mathbf{P_1} \sim \left(\frac{|\dot{R}(i\Delta t+t_0)|}{|\dot{R}(t_0)|}\right)^{\frac{1}{i\Delta t}} \) , the distribution of LE of residual component.  The KL divergence for recovery stability is given by:
{\small
\[
\frac{D_{\text{KL}}^r}{\log (R(t_0))}= -\sum_{i}  \mathbf{H}\left(P_{1}[x_i]\right)- P_1 [x_i]  e^{\gamma_1 \left[ \left(\frac{|\dot{R}(i\Delta t+t_0)|}{|\dot{R}(t_0)|}\right)^{\frac{1}{i\Delta t}}-0.5 \right]}
\]}
Where \(\mathbf{H(x)}=-x \ln (x)\) is the entropy.

\paragraph{Oscillation Index}
Oscillation Index is calculated by determining the KL divergence between the distribution of the LE (\( e^{\lambda_{IMF} }\)) of the oscillatory components (IMFs)  and the reversed Gompertz distribution. let \(\mathbf{P_2} \sim \left(\frac{|\dot{IMFs}(i\Delta t+t_0)|}{|\dot{IMFs}(t_0)|}\right)^{\frac{1}{i\Delta t}}\) , the distribution of LE of the residual component.  The KL divergence for oscillation stability can be given by:
{\small\[
D_{\text{KL}}^{imf}= -\sum_{i}  \mathbf{H}\left(P_{2}[x_i]\right)- P_2 [x_i]  e^{\gamma_2 \left[ \left(\frac{|\dot{IMFs}(i\Delta t+t_0)|}{|\dot{IMFs}(t_0)|}\right)^{\frac{1}{i\Delta t}}-1 \right]}
\]}
\subsubsection{Step 2: Tuning of \( \gamma_1 \)}
\begin{figure}
    \centering
    \includegraphics[width=0.8\linewidth]{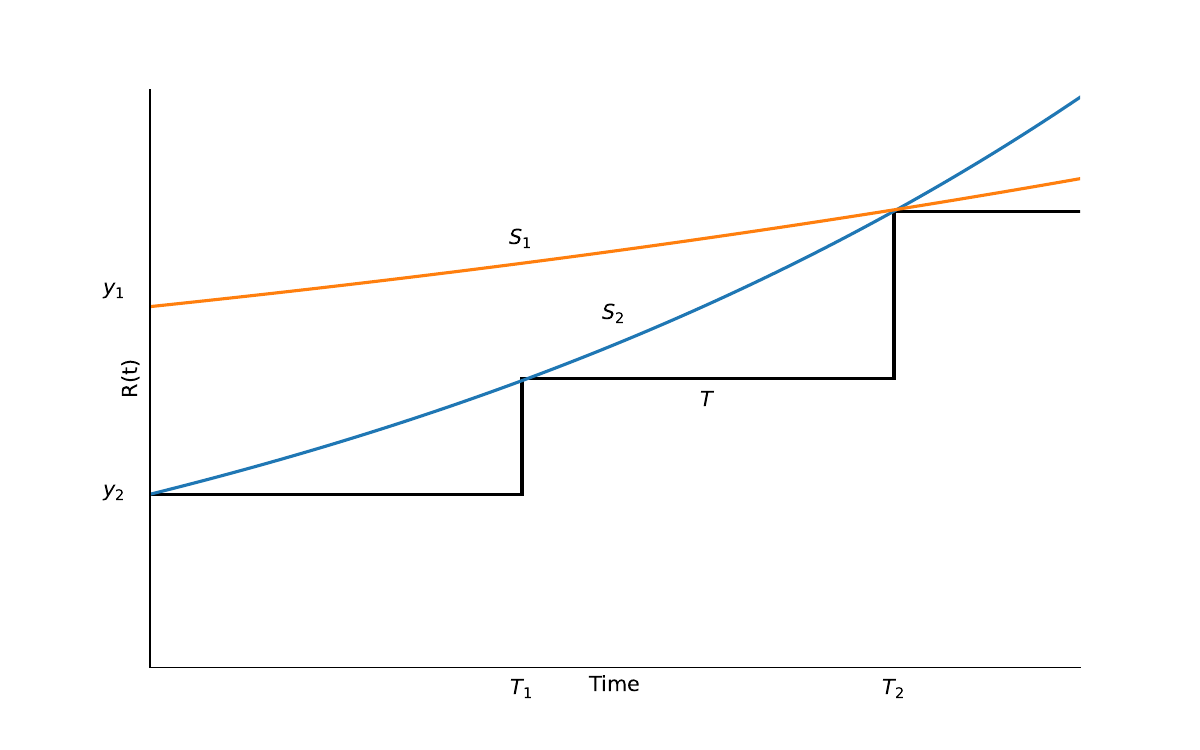}
\caption{Relationship between residual signals \(s_1\) and \(s_2\) and the tripping characteristic \(T\) of OEL or LVRT. 
}
    \label{fig:quasi-example}
\end{figure}

The parameter \( \gamma_1 \) should be selected as the maximum value that satisfy the following: 
\[
\max \gamma_1  \quad s.t, |D_{\text{KL}, s_1}^r - D_{\text{KL}, s_2}^r| < \epsilon
\]
The signals \( s_1 \) and \( s_2 \) are residuals of voltage signals and indicate the system at the verge of instability. Here, \( s_1 \) represents the slowest critical signal, while \( s_2 \) indicates the fastest critical signal, which not only triggers the OEL or LVRT but also destabilizes the system in the quasi-steady state. Figure \ref{fig:quasi-example} illustrates the concept of two interconnected signals (\( s_1, s_2 \)) that activate the tripping characteristic \( T \). Whether associated with OEL or LVRT, the concept is that any signal slower than \( s_1 \) starting from \( y_1 \), or any signal originating from \( y_2 \) and slower than \( s_2 \), will trigger the protection trip, leading to quasi-steady-state instability. Note that not all LVRT or OEL triggers lead to quasi-steady-state instability; however, we are interested here in the critical signals that do cause instability.

In the previous optimization problem, \( \epsilon \) denotes the index detection tolerance. If the index value of the measured signal exceeds the threshold by more than half \( \epsilon \), it indicates system instability; if it is less than the threshold by half \( \epsilon \), the system is considered stable. If the difference between the threshold and the index value is within half \( \epsilon \), the system is in a critical state. Where the threshold (\(D_{\text{critical}}^r\)) is defined as the average between \(D_{\text{KL}, s_1}^r\) and \( D_{\text{KL}, s_2}^r\). 

A lower \( D_{KL}^r\)index value indicates a higher degree of stability. Practically, there is no need to model the protection systems of OELs and LVRT to defined the threshold value (\(D_{\text{critical}}^r\)); it is sufficient to identify \( s_1 \) and \( s_2 \) through simulation. In the case of high penetration of single-phase motors, the recovery rate of the voltage profile is mainly influenced by the thermal relay settings of the air conditioner (AC).

\section{Simulation Results}

The simulation was carried out using the Nordic system at operation point A \cite{testSystems}, with the OEL modeled there. The thermal relay settings for AC motors, as well as the fault locations, durations, dynamic load percentages, and combinations of A, B, C, and D motors, were based on the cases provided in \cite{fundamental}. For the proposed index calculation, the simulation was run for 3 seconds after removing the fault to capture the voltage stability, while an extended simulation of 50 seconds was conducted to determine the quasi-steady-state stability due to the OEL triggering during delayed voltage recovery.

\begin{figure*}
    \centering
    \includegraphics[width=0.97\linewidth]{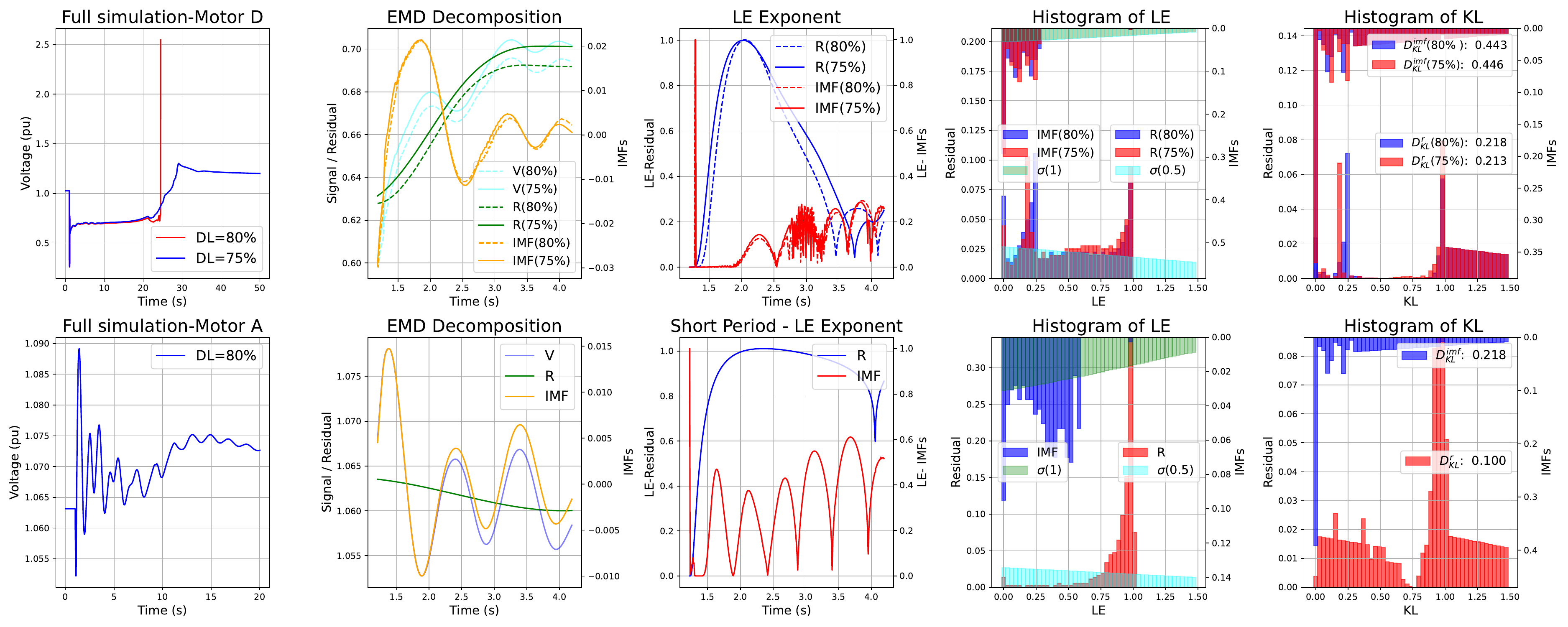}
 \caption{\textcolor{black}{Step-by-step calculation of the proposed stability index for two scenarios: the upper row illustrates a recovery case for Motor D with dynamic load levels of 80\% (unstable due to OEL triggering) and 75\% (stable), while the lower row presents an oscillatory stable case for Motor A. Each row includes the original voltage signal (V), EMD decomposition into residual (R) and IMFs, corresponding Lyapunov Exponents (LEs), probability distributions of LEs, and KL divergence calculations, demonstrating the index's ability to differentiate stability conditions based on both oscillatory and recovery components.}}
    \label{fig:OEL}
\end{figure*}

The first row of Figure \ref{fig:OEL} illustrates the step-by-step calculation of the proposed stability index for a system with high penetration of Motor D, comparing an unstable signal (dynamic load (DL) percentage = 80\%) and a close-to-stable signal (DL = 75\%). The system is simulated for 50 seconds to identify the critical signal based on the voltage trajectory, with instability occurring at DL = 80\%. The second column shows the EMD calculated over a short window (spanning over 3 seconds post fault), separating the residual and IMFs. The third column presents each component's LE, highlighting the stability characteristics. The fourth column shows the probability distribution of the residual (modeled with a shifted Gompertz function, \(\gamma_d = 0.5\)) and the IMFs (modeled with \(\gamma_d = 1\)). Finally, the fifth column displays the KL divergence between the residual/IMFs and the reference Gompertz function. The comparison of \(D_{KL}^r\) for the critical signal (DL = 80\%) and the close-to-stable signal (DL = 75\%) indicates that the proposed index effectively predicts system stability within a short period. The second row illustrates the same step-by-step process as in the first row but represents an oscillatory stable case for a system with high penetration of Motor A. The simulated voltage signal demonstrates stable oscillations, with a critical oscillation value of 1, while the measured value is 0.218, indicating that the system converges quickly after the fault. 

\textcolor{black}{While the instability occurs around 25 seconds, the proposed index effectively identifies unstable behavior within only 3 seconds after the fault, providing critical early warning. In addition, the index provides two separate measures: the oscillatory component and the residual component. In the real-scale system, the critical value for the residual component is determined offline (Section V), while the oscillatory component's threshold is set at 1. The index is then applied in real-time, calculated within 3 seconds post-fault, and compared against these predefined critical values.}

To illustrate the point about the degree of stability, we repeat the same process as in Fig. \ref{fig:OEL} for levels of dynamic load (DL) increasing from 0\%  to 100\%, with \(\gamma = 0.1\). The proposed index increases with an increase in DL share, reflecting the system's stability degree at each load level. For instance, at DL = 10\%, the index is approximately 0.05, indicating a highly stable system. As the DL increases to 20\%, the index rises to 0.15, reflecting a decrease in stability. When the DL reaches 60\%, the index further increases to 0.25, and at the critical load level of 80\%, it reaches the threshold value of 0.26, indicating instability. This commensurate increase demonstrates the index's capability to quantify the degree of stability across varying load conditions and detect instability early.

\section{Conclusion} 
In this study, we proposed a data-driven short-term voltage stability index to quantify the instability occurring because of the generator or IBR trip due to OEL or LVRT, respectively.
The index leverages EMD to decompose the voltage profile into IMFs and residuals. The LE is calculated over a short time window, and KL divergence is used to quantify the Lyapunov exponent for each decomposed component, with IMFs representing oscillations-induced stability and the residual indicating FIDVR-induced stability. Threshold values for each index are also computed to establish criteria for stability assessment. The proposed index is designed for real-time monitoring and offline analysis, providing timely and insightful stability assessment for modern power systems. 

\section*{Acknowledgment}

\small{This work was supported by the DOE through CyDERMS project (DOE CESER DE-FOA-0002503 award DE-CR0000040), the Power System Engineering Research Center (PSERC) and the National Science Foundation (NSF).}

\bibliographystyle{IEEEtran}  
\bibliography{ref}    
\end{document}